# Rotational Flows Generated by Microrobots Rotating Near Surfaces at Low Reynolds Number

Zhou Ye,[1,2,a)] Anirban Jana,[3,a)] and Metin Sitti[1,2,b)]

[1]*Max Planck Institute for Intelligent Systems, 70569 Stuttgart, Germany*

[2]*Department of Mechanical Engineering, Carnegie Mellon University, Pittsburgh, PA 15213, USA*

[3]*Pittsburgh Supercomputing Center, Carnegie Mellon University, Pittsburgh, PA 15213, USA*

Microrobots are untethered, mobile devices with characteristic lengths typically less than 1 mm, which have broad potential applications in biology, medicine, micromanufacturing, and microfluidics. Most of these applications require microrobots to operate in liquid environments. Motion of such robots inside a viscous liquid generates local flows around the robots, which can be used for tasks such as micromanipulation and micromixing. Therefore, knowledge of these induced flows is essential to the success of implementing the targeted tasks. In this study, we use numerical simulations to investigate the flow field induced by a single magnetic microrobot rotating with a constant angular speed about an axis perpendicular to an underlying surface. A parallel solver for steady Stokes flow equations based on the boundary-element method is used for simulating these flows. A simple transformation is introduced to extend the predictive capability of the solver to cases with small unsteadiness. Flows induced by four simple robot shapes are investigated: sphere, upright cylinder, horizontally-laid cylinder, and five-pointed star-shaped prism. Shapes with cross-sections that are axisymmetric about the rotation axis (sphere and upright cylinder) generate time-invariant flow fields, which could be useful for applications such as micromanipulation. Non-axisymmetric shapes (horizontally-laid cylinder and the star-shaped prism) induce significant unsteadiness inside the flow field, which could be desirable for applications such as micromixing. Furthermore, a slender horizontally-laid cylinder generates substantially three-dimensional flows, an added benefit for micromixing applications. The presence of nearby walls such as a bottom substrate or sidewalls has a retarding effect on the induced flows, which is quantified. Finally, we present the driving torque and power-consumption of these microrobots rotating in viscous liquids. The numerical modeling platform used in this work can enable future optimal microrobot designs for a given application requirement.

## I. INTRODUCTION

Microrobotics is a newly emerging field of robotics that has received significant attention from researchers in recent years.[1,2] Microrobots are untethered, mobile agents with characteristic lengths typically less than 1 mm.[1] Due to their small size and mobility, they can access enclosed spaces that are not easily accessible to conventional robots, such as inside of the human body and microfluidic lab-on-a-chip devices. Hence, microrobots are expected to find potential high impact applications in many areas such as biology, bioengineering, medicine, micromanufacturing, and microfluidics.[1-8]

Many of these applications require microrobots to navigate liquid environments, which, given sufficient liquid viscosity, reduces friction and adhesion between surfaces and sufficiently damps the fast dynamics of microrobots, leading to more

---

a) Z. Ye and A. Jana contributed equally to this work.

b) Corresponding author. Electronic mail: sitti@is.mpg.de.

stable and predictable robot motion control. Therefore, it is important to understand the interaction between the motion of a microrobot and the surrounding liquid to design and control appropriate microrobots for above applications.

Among the vast range of motions that microrobots can achieve, rotational motion has been especially attractive to researchers for the following reasons. First, it is easily implemented and highly controllable.[4,9-11] Second, rotational motion can be converted into other types of motions, such as translational motion, in both two-dimensions (2D) and three-dimensions (3D).[4,11,12] Finally, the rotational motion of the microrobots is critical to various applications, such as micromanipulation,[4,8,11,13,14] self-assembly,[15] and multi-robot control.[16] Due to the simplicity and significance of rotational motion, several works have already been carried out to investigate the interactions between rotating microrobots and the surrounding liquids. For example, Ye *et al.* explored the rotational flows generated by a spherical microrobot near a surface and how they can be used to achieve 2D non-contact micromanipulation.[4] Tung *et al.* also used the flow field generated by the rolling of a rod-shaped robot on a surface for non-contact micromanipulation.[14] Huang *et al.* studied the influence of robot shapes on the generation of microfluidic traps by rotation of microrobots with a cone-shaped head and a helical tail for 3D transportation of microobjects.[17] However, all these works focused on only specific applications without much generality or detailed understanding of robot-liquid interactions; hence, it is still necessary to conduct a more comprehensive investigation of robot-liquid interaction by computational fluid dynamics modeling.

In this work, we use numerical simulation tools to understand the low Reynolds number flows generated by a magnetic microrobot rotating at a constant angular speed near a surface about an axis normal to the surface. We primarily use a custom solver that is based on the boundary-element method to conduct the numerical simulations. This solver specifically solves the steady Stokes flow problem, completely neglecting both unsteady and nonlinear fluid inertia effects, by assuming that both the Reynolds number and the Womerseley number are strictly zero. We study how the robot shape affects the induced flow field, focusing on four typical robot shapes: sphere, upright cylinder, horizontally-laid cylinder, and five-point star. For the steady creeping flows generated by microrobots with circular symmetry about their axis of rotation (e.g., sphere and upright cylinder), the predictions of the steady Stokes flow code are very good. For unsteady flows generated by microrobots that do not possess circular symmetry about their axis of rotation, we show that the unsteady flow fields can also be accurately predicted by a simple transformation of the steady Stokes solution, if the unsteadiness is small. The unsteady flow fields possess some surprising characteristics, which we study in detail. The influence of gap size between the robot and the underlying surface on the induced flow field is also investigated. We further study how the presence of sidewalls affects the induced flows, because of their relevance to confined spaces inside the human body and microfluidic devices. Finally, we analyze the torque and power consumption requirements to rotate robots with different shapes in a viscous liquid. In summary, this study significantly improves our understanding of robot-liquid interaction for robots rotating near surfaces in the regime of very low Reynolds number (Re). The results are expected to be a useful guideline for designing microrobots for different applications.

## II. PROBLEM FORMULATION

In this study, we are interested in the influence of both robot shape and the presence of nearby walls to the flow field induced by the rotation of magnetic microrobots. The model we used for running numerical simulations is shown in Fig. 1. In the model, a cylindrical coordinate system ($r$, $\theta$, $z$) is defined with the origin located at the robot centroid. The robot always rotates about the $z$-axis with a rotational speed of $\omega$. The following geometric parameters are further defined: $R$ is the largest radial dimension of the horizontal cross-section of the robot body measured from the rotation axis (as shown in Fig. 1), $l$ is the height of the robot in the $z$-direction, and $g$ is the gap between the lowest point of the robot body and the bottom surface. The bottom surface is modeled as a circular plate with its center on the $z$-axis. In the presence of sidewalls, $d_{sw}$ represents the radial distance from the rotation axis to the sidewalls. The sidewalls have a height of $2.5R$, and surround the circumference of the bottom surface. The space between the robot and the sidewalls/bottom surface is assumed to be occupied by viscous Newtonian liquid of dynamic viscosity $\mu$ and density $\rho$. In all analysis of simulation results, geometric dimensions are normalized by $R$, while all velocities are normalized by $\omega R$. No-slip boundary conditions are applied on all solid surfaces.

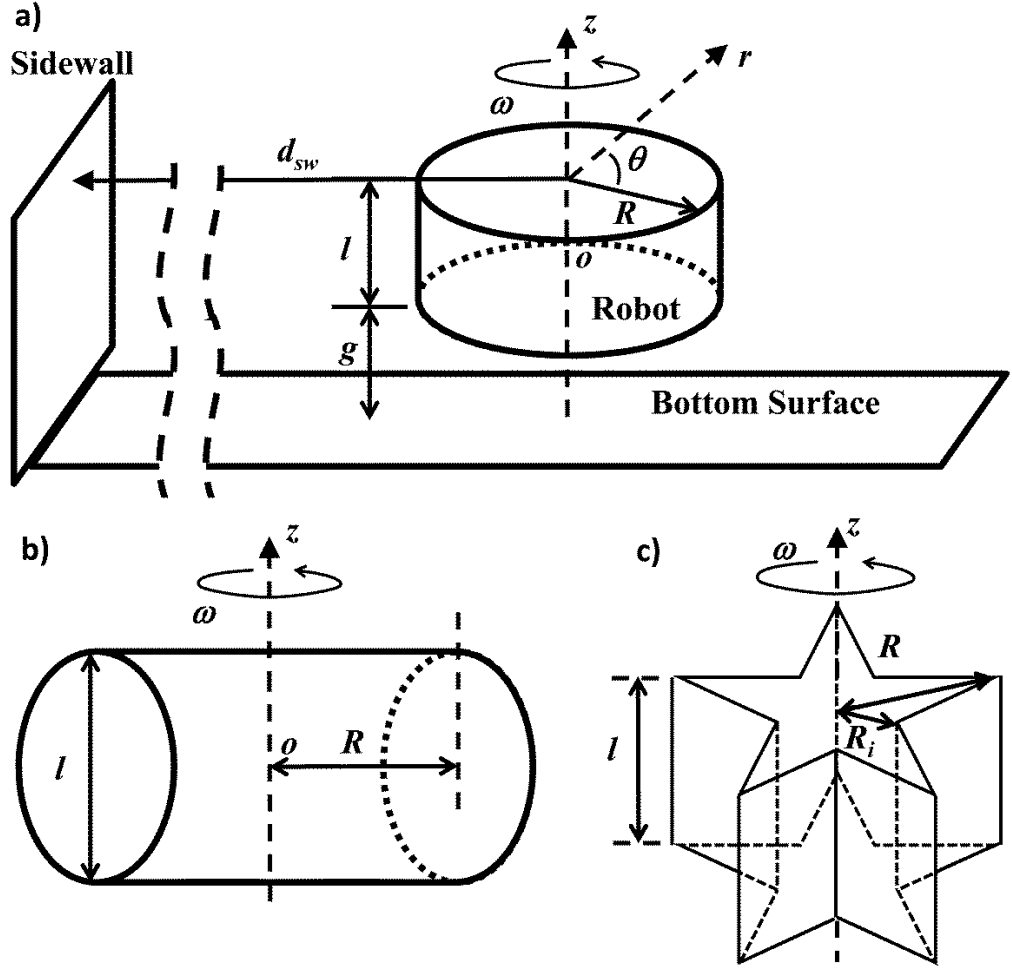

FIG. 1 Schematic of the models used in this study: (a) a robot body with axisymmetric cross-section about its rotation axis, (b) a horizontally-laid cylindrical robot body, and (c) a five-pointed star-shaped robot body with $R_i = 0.382R$. Neighboring surfaces are shown only in (a), but also exist in (b) and (c). In each case, the robot rotates around $z$-axis with a constant speed of $\omega$ at a distance of $g$ above a bottom surface. A cylindrical coordinate system $(r, \theta, z)$ is defined with the origin $o$ located at the centroid of the robot. The largest radial length of the robot body is denoted as $R$, while the robot's height in $z$-direction is denoted as $l$. In the presence of sidewalls, the distance from z-axis to sidewalls is denoted as $d_{sw}$.

## III. COMPUTATIONAL METHODOLOGY

### A. Boundary element method for Stokes flows

The computations are primarily performed using a boundary element method (BEM) code to solve the equations of conservation of mass (continuity) and linear momentum for steady Stokes flows. These equations, a simplification of the more general steady Navier-Stokes equations for Newtonian fluids whereby the nonlinear convective inertia term in the momentum equation can be neglected at very low Reynolds numbers, are as follows:

$$\mu\nabla^2 \mathbf{u} - \nabla p = 0$$
$$\nabla \cdot \mathbf{u} = 0 \qquad (1)$$

where $\mathbf{u}$ and $p$ are the fluid velocity and pressure fields, respectively, and $\nabla(\ldots) = \mathbf{e}_i \frac{\partial}{\partial x_i}(\ldots)$ is the gradient operator. Solutions of Stokes flows can be written in terms of Green's functions. For example, the velocity field $\mathbf{u} = u_j \mathbf{e}_j$ at an arbitrary location $\mathbf{x_0}$ in the flow can be written in terms of surface tractions on the bounding/immersed rigid surfaces as[18]

$$u_j(\mathbf{x}_0) = -\frac{1}{8\pi\mu}\int_S f_i(\mathbf{x}) G_{ij}(\mathbf{x}, \mathbf{x}_0) dS(\mathbf{x}) \qquad (2)$$

where $\mathbf{f}(\mathbf{x}) = f_i \mathbf{e}_i$ is the surface traction at the location $\mathbf{x} \in S$, $S$ being the union of all surfaces bounding or immersed in the flow, $G_{ij}$ is the following Green's function:

$$G_{ij} = \frac{\delta_{ij}}{d} + \frac{d_i d_j}{d^3}, \qquad (3)$$

$\delta_{ij}$ is the Kronecker delta and is equal to 1 if $i = j$ and 0 if $i \neq j$, $\mathbf{d}$ is equal to $\mathbf{x} - \mathbf{x}_0$, and $d$ is equal to $|\mathbf{d}|$. Note that we have used Einstein's notation for the tensors in the above equations.

In all our simulations, we have imposed velocities on rigid objects immersed in a viscous fluid, perhaps bounded by one or more stationary, rigid walls. Thus, if $\mathbf{x}_0$ is restricted to be on $S$, then the left hand side of Eq. (2) is known and the equation needs to be inverted to determine the unknown surface tractions $\mathbf{f}(\mathbf{x})$. Once $\mathbf{f}(\mathbf{x})$, $\mathbf{x} \in S$, is computed, velocites and all other flow quantities of interest can be determined everywhere in the flow.

In the BEM, the inversion of Eq. (2) is obtained by tessellating the boundary $S$ with a surface mesh consisting of, say, $N$ elements $S_k$, $k = 1, \ldots, N$, and evaluating the integral on the right hand side as a sum of integrals over each element. That is,

$$u_j(\mathbf{x}_0) = -\frac{1}{8\pi\mu}\sum_{k=1}^{N}\int_{S_k} f_i(\mathbf{x}) G_{ij}(\mathbf{x}, \mathbf{x}_0) dS(\mathbf{x}). \qquad (4)$$

By approximating each of these integrals using suitable numerical quadrature rules, a system of linear equations $[A]\{f\}=\{u\}$ is obtained, to be solved for the vector $\{f\}$ of unknown tractions at the nodes of the surface mesh. Note that the right hand vector $\{u\}$ consists of the known surface node velocities. Also note that whenever $\mathbf{x}_0$ is a node of element $S_k$, we find an integrable singularity that needs to be handled using a special method. Further details of the BEM for Stokes flows can be readily found in the literature.[18]

The boundary element code we are using is based on an open source code written in the programming language C.[19] We have extended the code to be able to simulate flows due to rotating rigid bodies. We have also parallelized the code using the PETSc library to enable fast simulation of complex systems.[20] Although the code is written for steady Stokes flows, it can also be used to study Stokes flows with small unsteadiness as long as the unsteady fluid inertia term $\mathbf{u}_{,t}$ is negligible compared to the viscous term. When the small unsteadiness assumption holds, the flow field depends on the instantaneous position and orientation of the rigid body (e.g., the microrobot) driving the flow. In this case, the steady state solver can be used to compute the instantaneous flow field at any particular instant, for instance, at $t = 0$. The instantaneous flow field at any other instant of time $t$ is then obtained by rotating the flow field at $t = 0$ together with the robot, using the transformation:

$$\mathbf{u}(r,\theta,t) = \mathbf{u}(r,\theta - \omega t, 0). \tag{5}$$

This is a powerful observation, as it enables one to predict the entire time-history of such flows with a single inversion of the discretized BEM equation, resulting in huge savings in computational cost.

Finally, we note the important advantages of the BEM over alternative discretization techniques such as the finite element method or the finite volume method. In the BEM, only the surfaces defining the flow domain and immersed objects need to be meshed. The fluid volume does not need to be meshed. Not only generating surface meshes is much easier than generating volume meshes, but the involved number of nodes and elements is also much less. Hence, the size of the linear system obtained via discretization is several times smaller as a result. Finally, large motion of rigid immersed bodies does not require mesh refinement when simulated using BEM, as would be the case if the fluid volume is needed to be meshed.

**B. Mesh independence study**

Mesh resolution has a significant impact on the accuracy of any numerical simulation. Therefore, we first conduct a mesh independence study to ensure that our BEM code provides satisfactory numerical accuracy before proceeding to the proposed problem. First, we simulate a rotating spherical microrobot and an underlying surface without any sidewalls. The gap between the robot and the surface is 1/10 of the robot radius $R$. From the simulation results, we extract the circumferential velocity in the flow field along radial directions on the plane $z = 0$ (where the robot centroid is located). It is known that the circumferential velocity decays with $r$ according to a power-law of the form:

$$v = \omega R \left(\frac{r}{R}\right)^b \tag{6}$$

For the case where there is no bottom surface for Stokes flow, the power-law exponent is $b = -2$.[21] However, due to the presence of the bottom surface, the flow velocity is expected to decay faster, resulting in a power-law exponent of $b < -2$.[4] As we refine the mesh further, the power-law exponent $b$ converges as shown in Fig. 2. A fit to the convergence data indicates a limit of -2.379 for $b$ as the surface element density goes to infinity, and with number of elements per normalized surface area $N_e$ = 750, the relative error in this value is already less than 0.1%. BEM simulations with such element density can easily run on normal desktop computers with 4-core CPU within tens of seconds, which is much more efficient than simulations requiring 3D body mesh such as finite-element-method or finite-volume-method. Furthermore, with increasing computational power, much higher element densities can be used for solving complex problems with high accuracy. In BEM simulations presented in the following sections, values of $N_e$ > 1500 were used for bottom surface and sidewalls, while values of $N_e$ > 3000 were used for robot surfaces. The total number of elements used in the BEM simulations is $2\times10^4$ or more, and the total number of nodes is at least $4\times10^4$. The resulting linear system matrix is dense with at least 14 billion double precision entries (matrix dimension is $3m\times3m$ for $m$ nodes) and requires 100-240 GB of memory for its storage alone. We solve such systems with the help of a supercomputer.[22]

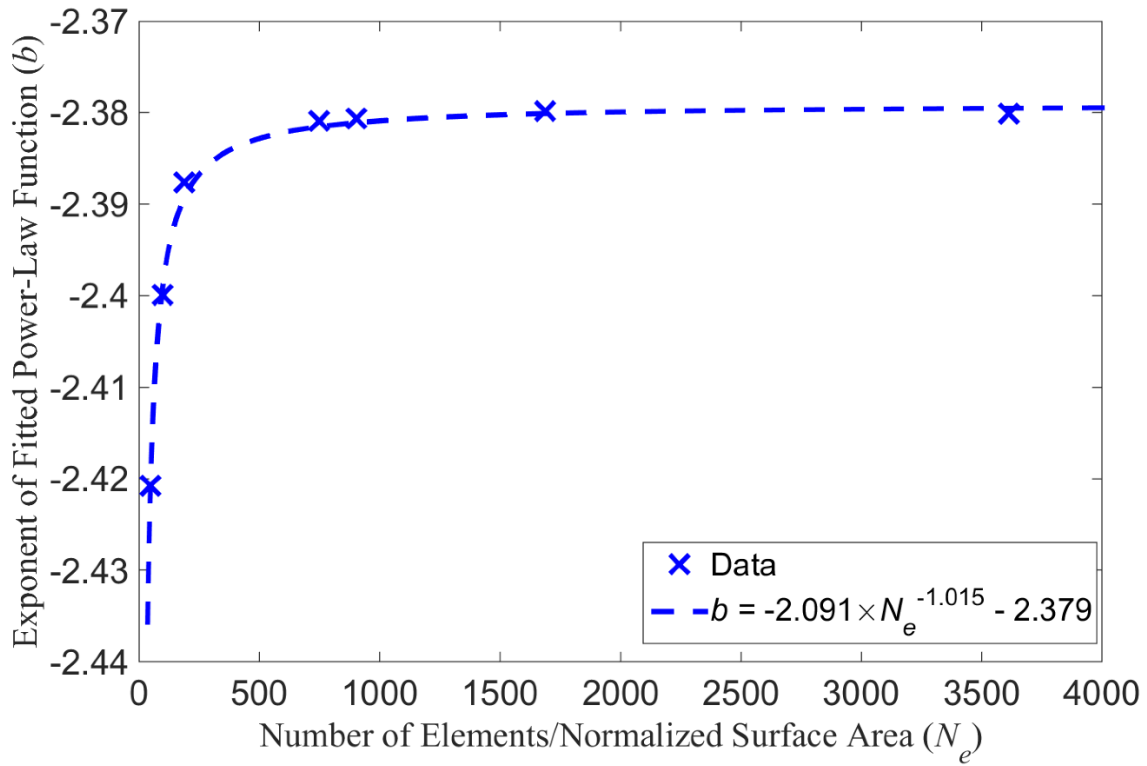

FIG. 2 Dependence of accuracy of numerical simulation on mesh resolution. The number $N_e$ plotted horizontally is a dimensionless measure of the surface mesh density. Here, "normalized surface area" is defined as the total surface area divided by the area of the robot surface. Plotted vertically is the extracted value of the exponent $b$ in the power law decay for $v(r)$.

## C. Validation of the BEM code for steady flows generated by a rotating microrobot

Microrobots possessing axisymmetry about their axes of rotation, such as the sphere and upright cylinder in our study, generate steady flows. To verify the BEM code and assess the validity of the Re = 0 assumption of Stokes flows for microrobot applications, we compare the solution of our Stokes flow BEM code to that of COMSOL for a steady flow case. COMSOL is a commercial finite element package and is capable of solving the full Navier-Stokes equations for fluid flows for small as well as large Re. In this validation study, we use a model of a spherical microrobot rotating inside an open-top container consisting of four sidewalls and a bottom surface, which is filled with a viscous liquid. The distance from robot to the sidewalls, $d_{sw}$, is 8.3$R$, and the gap $g$ is 0.1$R$. In the COMSOL model, the viscous liquid medium in which the robot rotates is chosen to have a dynamic viscosity of $1\times10^{-3}$ Pa·s, and a density of $1\times10^3$ kg/m$^3$, corresponding to Re = 0.2.[4] We find that the COMSOL model predicts the power-law exponent $b$ for the decay of circumferential velocity with radial distance to be -2.430, while our BEM code predicts that $b$ = -2.436 for this case. Thus, the agreement between COMSOL and our BEM code is excellent even at a moderate Re of 0.2. The relative difference between the two results is less than 0.3%. In practice, microrobotic applications will typically have a Re at 0.001 – 0.1, further improving the BEM predictions (velocity errors in Stokes solution decrease as $O(\text{Re})$[21]). The slight discrepancies in the $b$ value could arise from the fact that COMSOL solved for the full Navier-Stokes equations, which include all inertial effects, while the custom BEM code only accounted for non-inertial effects. Note that the value of $b$ is even more negative in this case than when only a bottom surface is present, implying that the presence of sidewalls has an additional retarding effect that causes $v(r)$ to decay even faster.

## D. Validation of the BEM code for unsteady flows generated by a rotating microrobot

The BEM code used in this work only solves for steady Stokes flow problems, while the rotation of non-axisymmetric robots induces unsteadiness in the flows generated by such motion even under Stokes flow condition. However, as argued in Section IIIA, as long as the unsteadiness is small, it is expected that the steady Stokes solution obtained by our current BEM code, together with the transformation given by Eq. (5), should provide a good approximation of these unsteady flows. In this section, we examine to what extent our above hypothesis is true, by comparing our BEM predictions with that of COMSOL for the flow generated by a slender, horizontally-laid cylindrical rod rotating at the center of an upright cylindrical container without top or bottom lids. The rod has a length of 20 μm and a diameter of 2 μm. The container has a diameter 5 times the length of the rod, and is filled with water. In COMSOL, the unsteady Stokes equations are solved for the above case. Then, we compare the time history of the *y*-velocity of the flow at a given spatial location (obtained by combining our BEM solution with the transformation given by Eq. (5)) to that obtained from COMSOL, at a given spatial location, over a full rotation period of the rod. We have performed this comparison for three different rotation speeds of 5 Hz, 50 Hz, and 500 Hz. For the sake of brevity, the comparison is shown only for the latter two rotation speeds in Fig. 3. We observe that at the two lower rotation speeds (5 Hz and 50 Hz), the BEM and COMSOL predictions match well, demonstrating that at these rotation speeds, the unsteadiness is small enough to permit us to combine the steady flow solutions from our BEM code with Eq. (5) to accurately predict the time histories of the unsteady flows. However, at the higher rotation speed of 500 Hz, the unsteadiness becomes significant, creating a large discrepancy between the two predictions. In this case, the COMSOL prediction is still correct, but the approximation using our current BEM code is not. In general, the maximum rotation speed up to which the transformed steady BEM result gives good prediction of the unsteady flow also depends on the aspect ratio of the rod. The detail analysis on the degree of unsteadiness in the flows is out of the scope of this work and would be a future work.

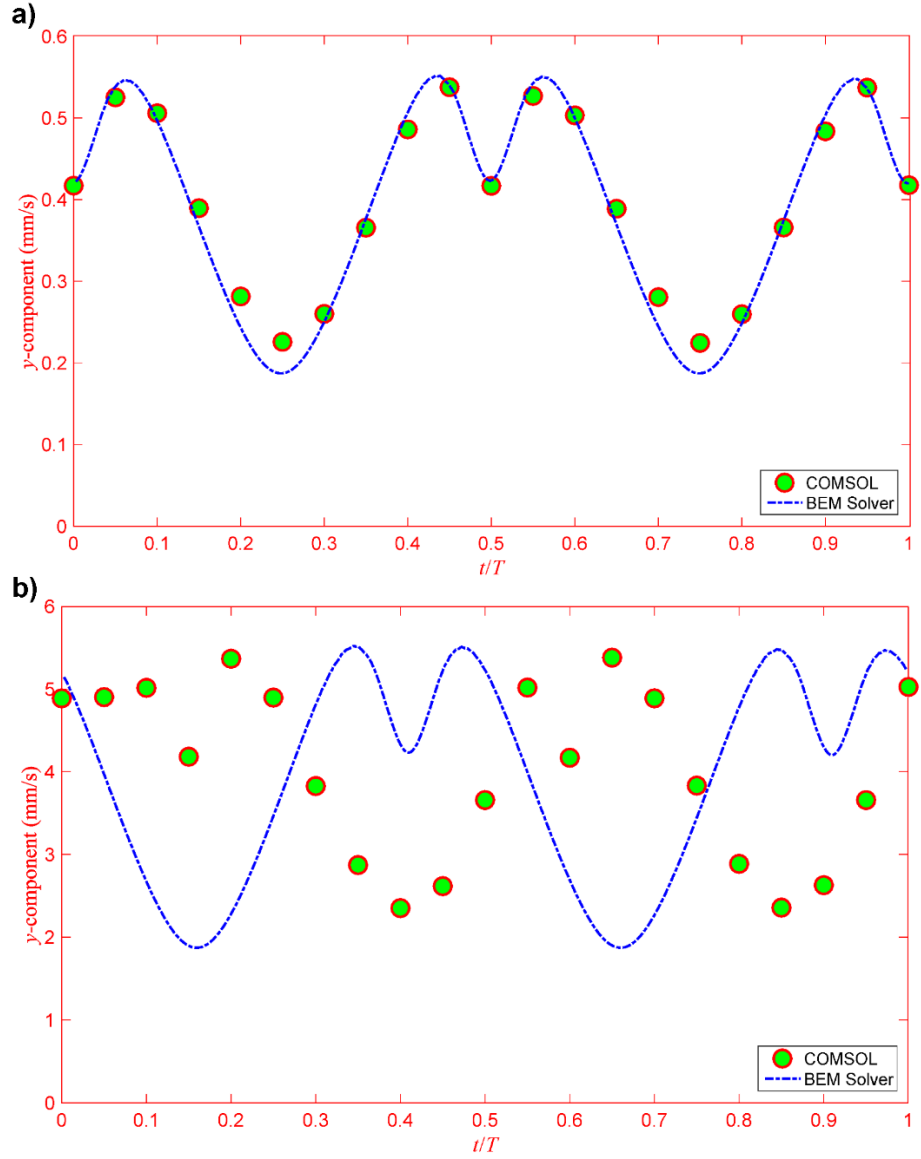

FIG. 3 Effect of flow unsteadiness on the validity of our BEM code for the prediction of time-varying flows. The solution from our steady BEM solver combined with the transformation given by Eq. (5), and the unsteady Stokes solution from COMSOL are compared for the case of a slender rod rotating in water. The comparison is performed for the *y*-velocity of the fluid at the location $(r, \theta, z) = (1.1R, 0, 0)$ over one complete rotation period $T$. The comparison is shown for two rotation speeds: a) 50 Hz, and b) 500 Hz.

## IV. RESULTS AND DISCUSSIONS

### A. Effect of robot shape

In the first set of simulations, our aim is to study how the robot's shape affects the flow field induced by its rotation. In this set of simulations, a robot with a particular shape rotates above a bottom surface with a finite gap $g$. Four basic robot shapes are investigated: sphere, upright cylinder, horizontally-laid cylinder, and a prism with a cross-section of a regular five-pointed star. The following parameters are kept constant for all the simulations of this set: maximum radial dimension $R$ of the robot, height of the robot $l = 2R$, gap between the robot and the underlying surface $g = 0.1R$, and the rotational speed $\omega$. Finally, for the bottom surface, a circle of radius $8R$ is used. No other boundary walls except the bottom surface are present. The flow velocity data are computed on a rectangular grid encapsulating the robot and are used for all of the subsequent analyses. The streamlines depicting the induced flow field around each robot shape are presented in Fig. 4. The streamlines are generated in MATLAB using the instantaneous velocity field data extracted from the simulations. First, we note that in all cases, there exist no net out-of-plane or radially outgoing flows in the flow fields induced by the rotating robots under Stokes flow conditions. As a consequence, all the streamlines are closed curves. The absence of axial suction and radial ejection of fluid driven by rotation, often-called "centrifugal pumping", is due to the assumption of Stokes flow conditions. The full Navier-Stokes equations predict the centrifugal pumping phenomenon to occur at even very low Reynolds numbers. However, the strength of the centrifugal pumping is weak at low Reynolds numbers and scales as $O(\text{Re})$. For example, Liu and Prosperetti[23] show that the maximum axial flow velocity towards a rotating sphere scales approximately as $\text{Re}/75$, for small Re. Hence for microrobotic applications, the inability of Stokes flow to capture centrifugal pumping phenomenon is expected to result in only a small error, e.g., in the radial force balance for trapped objects in micromanipulation.

For the sphere and the upright cylinder, the streamlines are circular as expected. For the five-pointed star, the streamlines tend to closely conform to the robot shape in the vicinity of the robot, but smoothen to pentagon with rounded corners at a distance. The smoothening is more pronounced near and beyond the top of the robot. Finally, for the horizontal cylinder, the streamline shape in the vicinity of the robot bears a complex relationship to the robot shape. The streamline shapes resemble rectangles but with rounded corners and wavy edges, and rotated by 45° relative to the rectangular horizontal cross-sections of the robot. The streamlines come closest to the robot near its two edges, and move away from the robot over both the flat and curved faces. For the sphere and the upright cylinder, the streamlines lie completely in planes parallel to $z = 0$. For the horizontal cylinder with unity aspect ratio and the star, the streamlines also almost completely lie on planes parallel to $z = 0$, with some very small out-of-plane excursions. This is not the case though for long and slender horizontal cylinders, as discussed later in Section IVE.

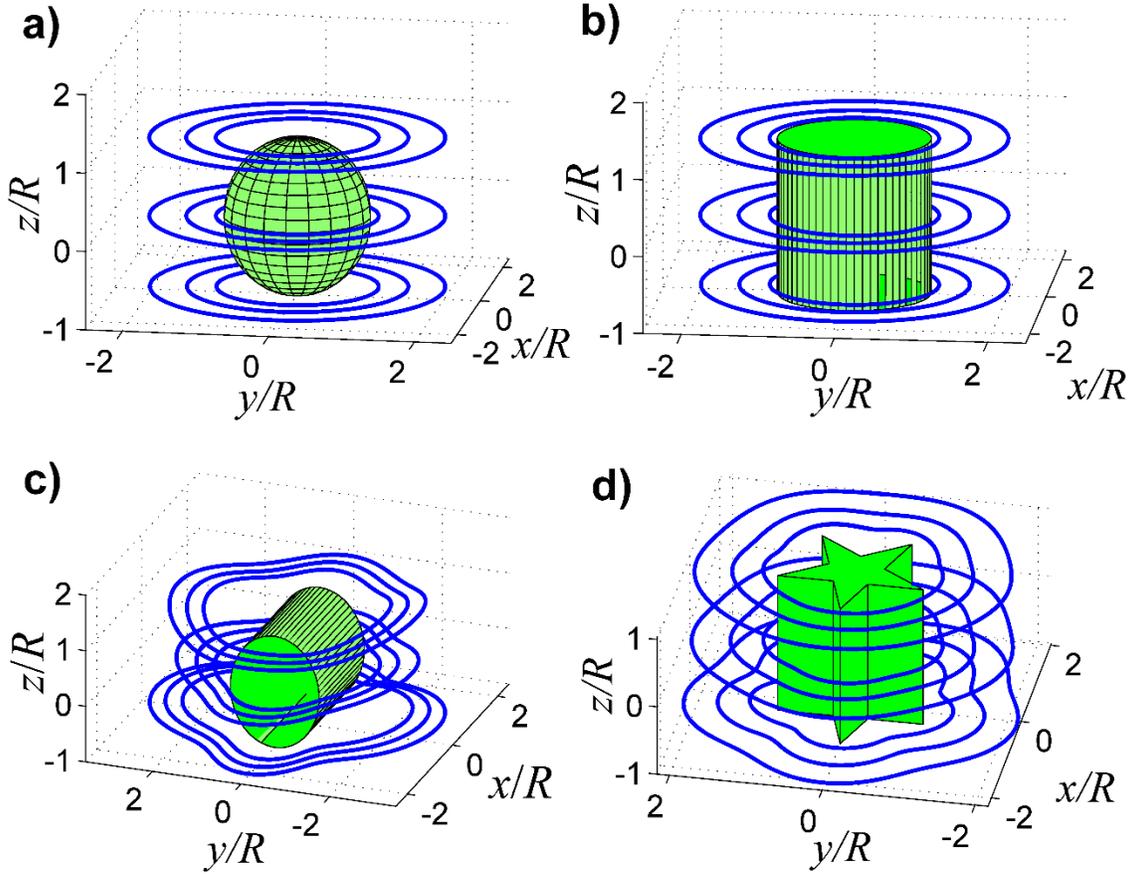

FIG. 4 Streamlines of the induced flows around robots rotating about $z$-axis with different shapes. The blue solid lines indicate the streamlines while the green solid features represent the robot body. a) Robot with a sphere shape. b) Upright cylinder. c) Horizontally-laid cylinder. d) Five-pointed star. Streamlines are oriented clockwise or counter-clockwise according as the sense of robot's rotation about $z$-axis is clockwise or counter-clockwise.

To gain a more quantitative insight into the nature of the induced flow fields for the different robot shapes, we next look at how the flow velocity behaves as a function of $r$ and $z$. We look at the circumferential velocity component $v$, which is the dominant velocity component in the flow field (and in fact, the only component of velocity predicted by Stokes flow for axisymmetric shapes). In Fig. 5, the circumferential flow velocities, normalized by $\omega R$, are plotted along a radial direction on the plane at $z = 0$, where the robots' centroids are located. For the horizontally-laid cylindrical and star-shaped robots, for which the flow velocities are time-dependent, the velocity data plotted have been averaged over one full rotation cycle. We see that for all robot shapes, $v$ is maximum at $r/R = 1$, and decays monotonically as $r$ increases. For both axisymmetric shapes (sphere and upright cylinder), $v/\omega R$ starts off at the peak value of unity at $r/R = 1$. This is because the circle $r = R$, $z = 0$ lies entirely on the robot surface and the circumferential velocity on the surface of the two axisymmetric robots at $z = 0$ is constant and equal to $\omega R$ regardless of $\theta$. For the two non-axisymmetric shapes (horizontal cylinder and star), the peak value at $r/R = 1$ is less than unity. That is because any spatial point on the circle $r = R$, $z = 0$ can be either on the robot surface or at other times in the fluid radially away from robot, and hence has circumferential velocity = $\omega R$ only for a few instants in each rotation cycle. Since the flow velocities decay away from the robot surface, for most of the rotation cycle, the circumferential flow velocity at such points is less than $\omega R$, and hence the time-averaged normalized value is less than 1. Similar to the sphere case, we also fit the velocity data for other shapes and obtain the following relationships: the fitted function for upright cylinder is $\frac{v}{\omega R} = 3.382 e^{-1.219(r/R)}$, for horizontal cylinder is $\frac{v}{\omega R} = 1.988 e^{-1.079(r/R)}$, and for the star-shape is $\frac{v}{\omega R} = 2.983 e^{-1.318(r/R)}$. It is seen that in fact the normalized average circumferential flow velocity, induced by an axisymmetric shape, is always larger than that of a non-axisymmetric shape at any given radial location $r$. The difference is especially pronounced in the vicinity of the robot. When $r/R$ is within the range 1.5 – 3, which is the typical range of trapping distances in micromanipulation, the normalized average circumferential flow velocity induced by a non-axisymmetric shape is only 70% – 80% of that induced by an axisymmetric shape. Thus, axisymmetric robots will in general be more efficient in micromanipulation tasks. Among the two axisymmetric shapes, the upright cylinder induces stronger fluid rotation everywhere when compared to the sphere. Hence in terms of the hydrodynamic force exerted on target objects to be manipulated, the upright cylinder is the best choice. However, the sphere can be easily repositioned in the workspace by tilting its rotation axis and initiating a rolling motion simultaneously with the main rotation.[4] Thus, in micromanipulation applications where robot repositioning is necessary, the sphere may be preferred over the upright cylinder. However, it should be noted that in the case where the rotation axis of the cylindrical robot is parallel to the underlying surface, the robot could also be precisely repositioned while generating the flow field for manipulation.[14]

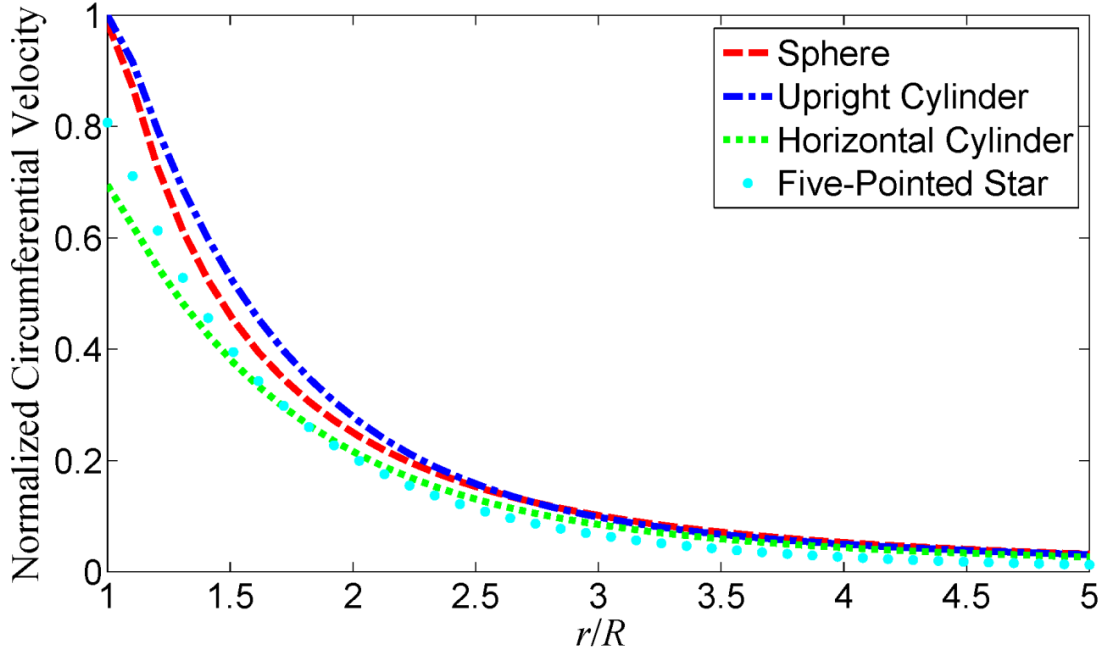

FIG. 5 Variation of circumferential flow velocity along radial direction $r$ at the height of robot centroid ($z = 0$) for different robot shapes. The circumferential velocities for non-axisymmetric shapes were averaged over one full rotation cycle at corresponding locations. All velocities were normalized by $\omega R$.

Next, the velocity distributions along the $z$-direction are investigated. The variation in circumferential velocity along the $z$-direction at $r/R = 1.1$ for all the four robot shapes are plotted in Fig. 6. As before, for the non-axisymmetric shapes, the averages over a rotation cycle are plotted. It is observed again that the upright cylindrical shape generates the strongest fluid rotation among all shapes tested. The presence of the bottom surface makes all the velocity profiles asymmetric about $z = 0$. It is also observed that for the three robots that have a constant $R$ value along their height $l$ (the two cylinders and the star), the flow velocity $v$ is virtually constant for a significant portion of the robot height $l$. To characterize this feature more quantitatively, we define a new term, the Strong Rotation Layer (SRL), defined as the range of $z$-values within which the flow speed $v$ is at least 70% of its maximum. As shown in Fig. 6, the upright cylindrical shape generates the thickest SRL, while the spherical shape generates the narrowest, with the thickness less than half of that for the upright cylinder. Also, we see that the maximum speed occurs slightly above the plane of robot centroid ($z = 0$) for the upright and horizontal cylinder and five-pointed star shapes, due to the presence of the bottom surface. For the sphere, however, the shift in the location of the maximum from the $z = 0$ plane is negligible. In general, these results reiterate the observation that in terms of the strength of the hydrodynamic force exerted on trapped objects, the upright cylinder has the best performance. Also, since the velocity *v* increases the most rapidly and becomes significant even close to the bottom surface, the upright cylindrical robot should be able to trap and manipulate even very small objects lying on the surface.

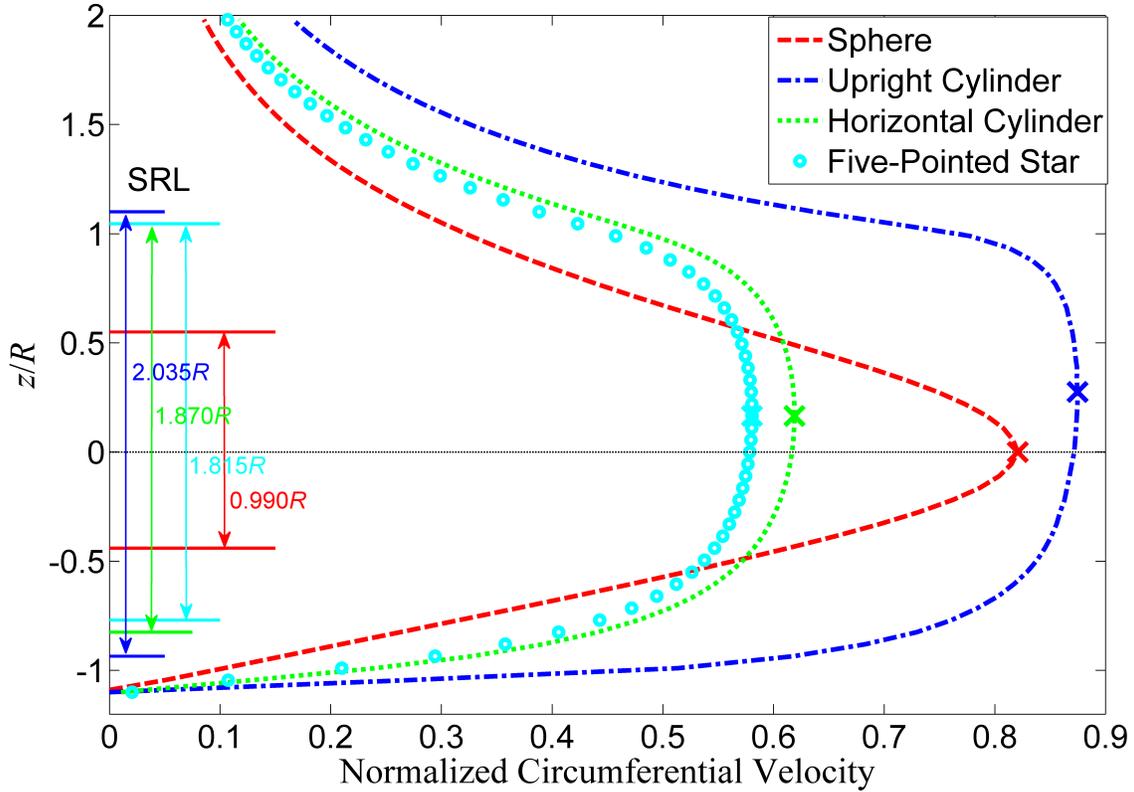

FIG. 6 Variation of circumferential velocity along *z*-direction at a given radial distance in the flow field by different robots. The velocity data were taken at a distance of $r/R = 1.1$, and normalized by $\omega R$. The robot centroid locates at $z/R = 0$ on the *y*-axis in the figure. SRL stands for strong rotation layer, the region of which is defined as the range along *z*-axis within which the circumferential flow speed is at least 70% of its maximum ($|v| \geq 0.7|v|_{max}$) at the given distance in the flow field. Colors corresponds to different robot shapes. The "x" marks on the velocity profiles indicate the locations of the peak velocity values on the profile.

### B. Effect of gap between robot and bottom surface

Due to the no-slip condition, the bottom surface acts to retard any nearby flow. This retarding effect thus acts on the rotational flows induced by the rotating robot and depends on the size of the gap *g* between it and the robot. It is expected that a smaller gap would result in a stronger retarding effect. As a result, the radial decay of the flow velocity on a given horizontal plane should be faster for a smaller gap size, and the exponent *b* of the decay power law (Eq. (2)) should be more negative. In Section IIIB, we have already seen that for an infinitely large gap *g* (i.e., no bottom surface), $b = -2$, and for $g = 0.1R$, $b = -2.379$. In this section, we present in greater detail the dependence of the exponent *b* on different gap sizes. The results in this section are based on a set of simulations of a spherical robot rotating above a bottom surface with no sidewalls. All other parameters except the gap size $g$ are kept constant throughout this set of simulations.

The variation of the exponent $b$ of the fitted power law decay of $v(r)$ with the normalized gap size $g/R$ is shown in Fig. 7. We see that the dependence of the exponent $b$ on the normalized gap size $g/R$ is also exponential. At zero gap, $b = -2.428$. As gap size $g$ increases, the exponent $b$ increases monotonically and asymptotically approaches a limit for large gaps. The limit of the exponent $b$ indicated by the fit as $g \to \infty$ is $b = -2.039$, which corresponds to the case of a sphere spinning in free space under Stokes flow conditions. The above limiting value is very close to the analytically computed limit of $b = -2$.[20] The discrepancy between these two values is less than 2%, and is probably due to small errors from simulation and fitting.

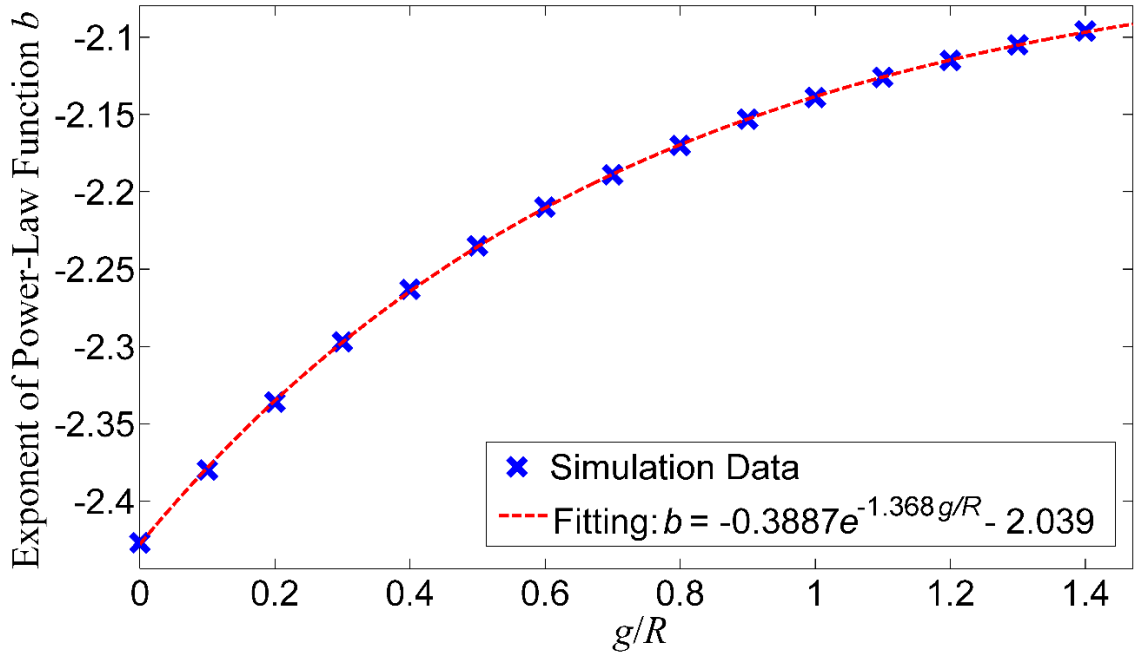

FIG. 7 Exponents of power-law fitting function for circumferential flow velocity along radial direction at the height of robot centroid ($z = 0$) for a spherical robot rotating above the bottom surface with different gap size $g$.

## C. Influence of sidewalls

The presence of sidewalls also affects the induced rotational flows. Simulations are carried out to study this effect, using a spherical microrobot rotating above a bottom surface at a gap size $g = 0.1R$, and sidewalls at a distance of $d_{sw}$ from the $z$-axis. All the parameters except for $d_{sw}$ are kept constant in this set of simulations.

Similar to Section IVB, we look at the circumferential velocity as a function of the radial distance $r$ on the plane $z = 0$ (on which the robot's centroid is located). The results are plotted in Fig. 8. As mentioned in Section IIIC, sidewalls tend to slow down the flow induced by the robot. In general, the closer the sidewalls to the robot, the lower the flow velocity throughout the space. However, this reduction in flow velocity is only significant very close to the sidewalls. The reduction is 20% or greater at locations within 1 robot radius from the sidewalls.

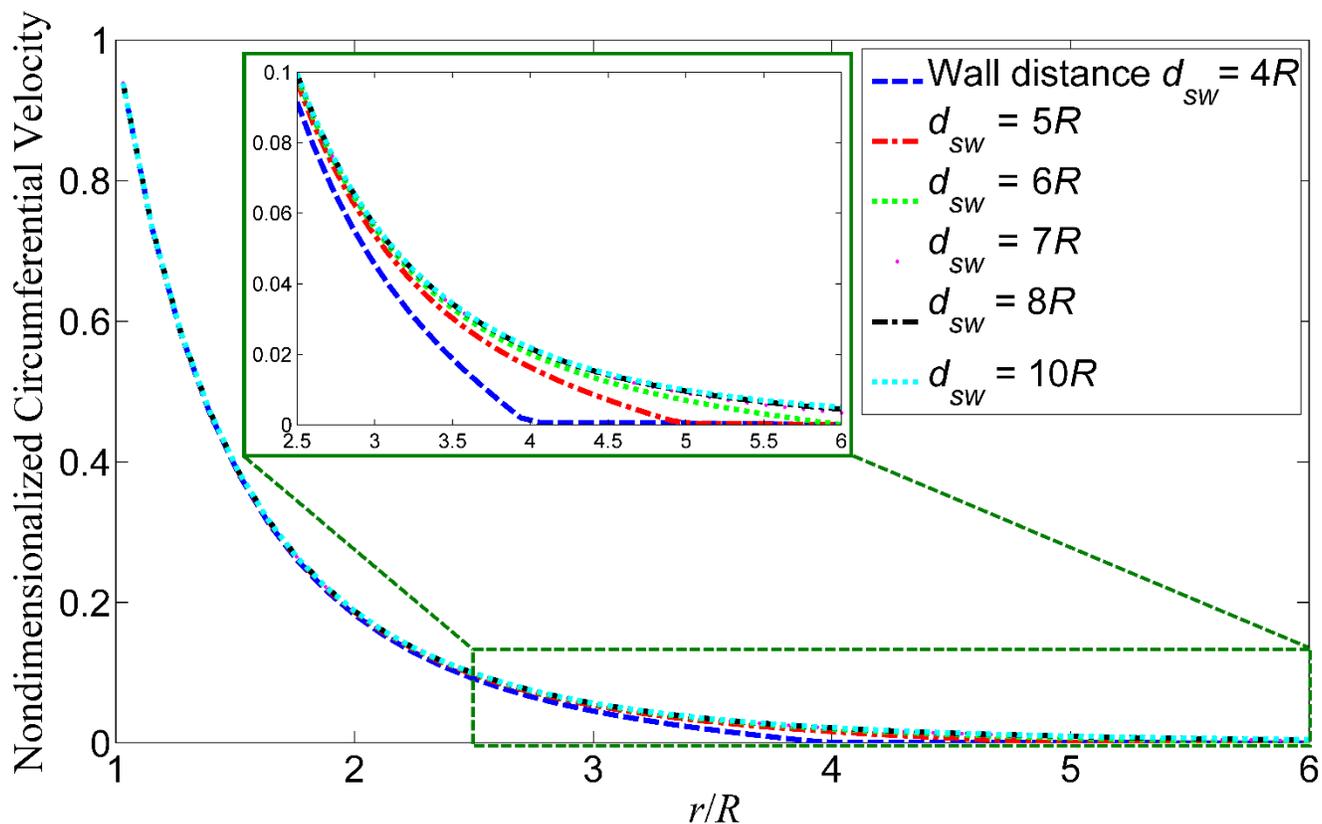

FIG. 8 Influence of sidewalls on the circumferential flow velocity profiles along the radial direction $r$ at $z = 0$ in a flow field induced by a spherical robot.

## D. Flow unsteadiness generated by non-axisymmetric robots

Robots that have a non-axisymmetric shape about the rotation axis, e.g., the horizontally-laid cylinder and the five-pointed star in this study, induce unsteady flows due to their rotation. Section IIID provided a glimpse of the unsteady character of these flows. In this section, we describe the unsteadiness in more detail. Figure 9 shows the variations of flow velocity magnitudes over one period of rotation for the horizontally-laid cylinder and the five-pointed star respectively. Time histories at 8 different radial locations on the $z = 0$ plane are shown. At any given location, the flow velocity magnitude is seen to oscillate about a mean value for both the robots. As described in detail in Section IVA, the mean flow velocities decay with increasing radial distance. The flow oscillation amplitudes also decay with increasing radial distance, i.e., in both cases the unsteadiness becomes smooth at greater distance from the robot. The time-histories for the five-pointed star-shaped robot are periodic with a fundamental frequency $f_1$ equal to 5 times the rotation speed $\omega$ (in Hz), as expected from the star's geometric periodicity $N_g = 5$. The temporal oscillations are not far from sinusoidal, which indicates the higher harmonics are weak compared to the fundamental. The deviation from the sinusoidal is most prominent close to the robot. For the horizontally-laid cylinder, the geometric periodicity $N_g = 2$ suggests that the fundamental frequency be $f_1 = 2\omega$. This is technically true, and more clearly discernible at a large distance from the robot (two wide and two narrow peaks per rotation period for $r \geq 2.25R$). But closer to the robot ($r \leq 2R$), the fundamental frequency appears to be $4\omega$, which can be understood by noting that any given spatial location sees an edge pass by 4 times in one rotation period. The waveforms near to the horizontally-laid cylinder are much more complicated than those for star, indicating the substantial presence of higher harmonics. Fourier analysis shows that typically, the 2nd harmonic $f_2 = 4\omega$ is the most dominant (so dominant close to the robot that it appears to be the fundamental), and the first 4 harmonics (up to $f_4 = 8\omega$) contribute significantly to the overall waveform (the amplitude is at least 10% of the most dominant harmonic). As an exception, at $r = 1.75R$, the 4th harmonic ($f_4 = 8\omega$) is the most dominant, and all harmonics up to the 8th ($f_8 = 16\omega$) significantly contribute. In summary, the unsteadiness in the flow generated by the horizontal cylinder is more complicated than that generated by the five-pointed star.

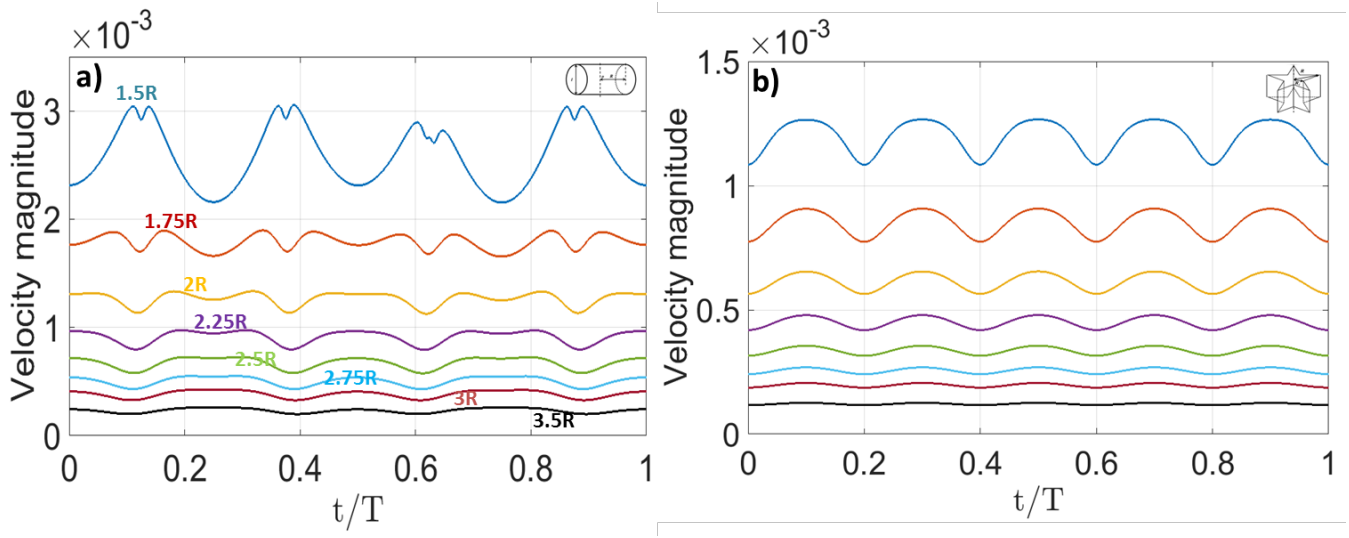

FIG. 9 Time histories of the unsteady flow velocity magnitudes induced by the rotation of a) the horizontally laid cylindrical robot with unity aspect ratio, and b) the five-pointed star-shaped robot. The time histories correspond to 8 different locations along a radial line on the $z = 0$ plane: $r = 1.5R$, $1.75R$, $2R$, $2.25R$, $2.5R$, $2.75R$, $3R$ and $3.5R$.

**E. Three-dimensionality of the flows induced by horizontally-laid cylindrical robot**

The flows generated by the horizontally-laid cylindrical robot are additionally complicated by the presence of strong 3D effects, especially when the cylinder is long and slender. This distinguishes the horizontal cylinder from the other three shapes examined, which have primarily 2D shapes. Figure 10 shows an instantaneous snapshot of the velocity field on the $z = 0$ plane generated by a long and slender horizontal cylinder rotating $g = 0.1R$ above a bottom surface. To illustrate the 3D nature of this flow, the velocity vectors are colored by the local value of the out-of-plane or z-component of the velocity $w$. Although the out-of-plane component is approximately 10 times smaller than the in-plane components of the velocity, this component is critical to creating the 3D aspects of the flow. It is observed that significant out-of-plane flow exists on the $z = 0$ plane (even though it is a plane of symmetry for the cylinder itself) in four regions near the edges of the curved cylinder surface. In front of the two leading curved surfaces, the flow is directed out of the plane along the $+z$-axis, whereas behind the two trailing curved surfaces, the flow is directed into the plane along the $-z$-axis. We believe that this phenomenon is due to the constraints on the liquid by the moving curved surface of the cylinder on one side, and the stationary bottom surface on the other side. The portions of the leading curved surfaces above the $z = 0$ plane push the liquid in front of it forward and upward as it moves. On the other hand, portions of the leading curved surfaces below the $z = 0$ plane push the liquid in front of it forward and downward, but due to the presence of the bottom surface, this body of fluid is also deflected upwards. The combined effect is a net upward component of the flow in the region in front of the leading curved surfaces of the cylinder. The reverse situation occurs behind the trailing curved surfaces whose motion generates suction or low pressure in these two regions. Due to the presence of the bottom surface, the liquid can only flow downward from above the cylinder to fill in this void.

We note that in the case of the rotating horizontal cylinder with no bottom surface, the flow field on the $z = 0$ plane is 2D in nature due to up-down symmetry. However, both above and below the $z = 0$ plane, the flow still possesses out flow components due to the motion of the curved cylinder surface. Thus the global flow field exhibits significant three-dimensionality even in the absence of a bottom surface.

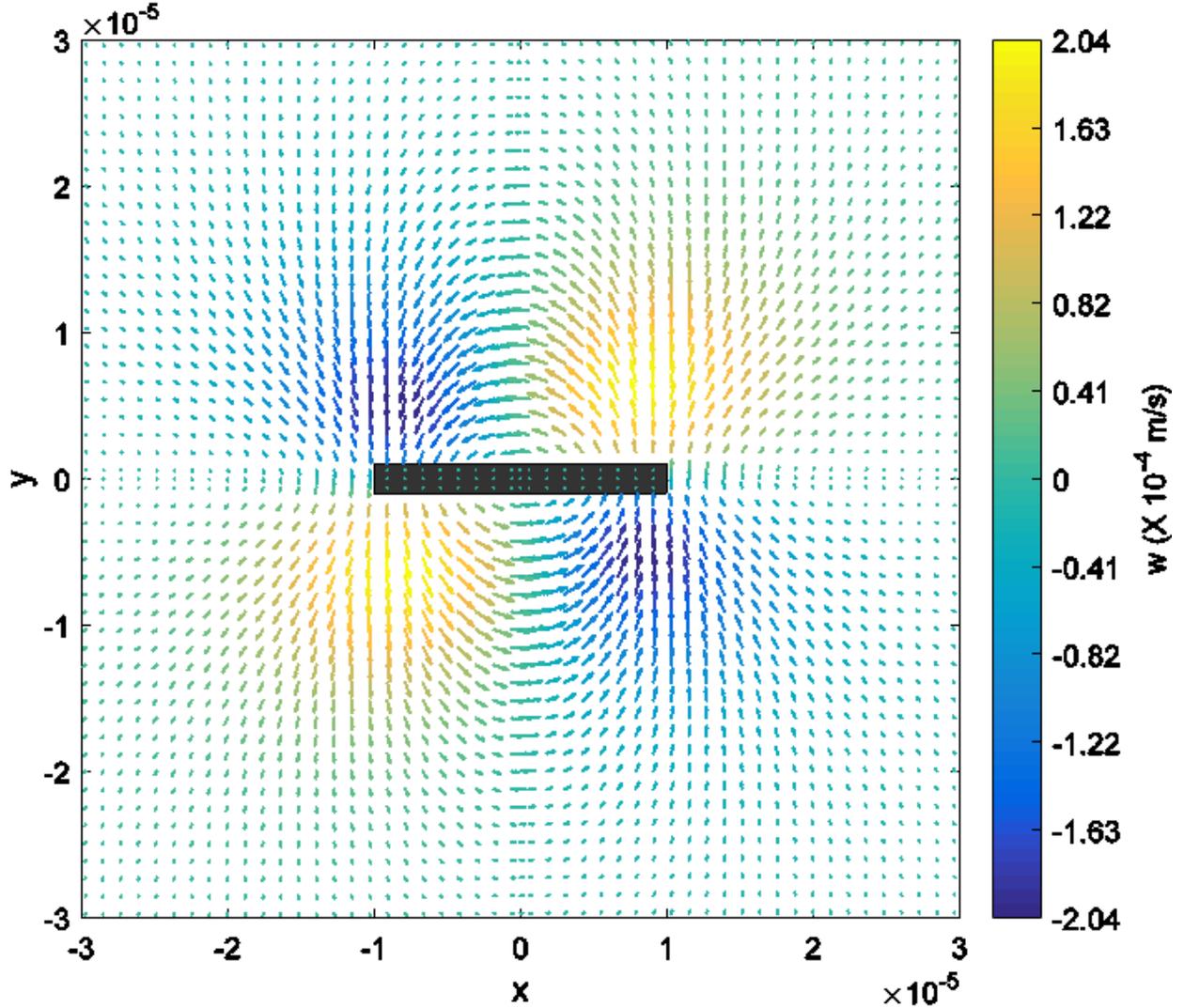

FIG. 10 Instantaneous velocity field on the $z = 0$ plane generated by a horizontal cylinder with aspect ratio $\eta = 0.1$ rotating counterclockwise with a gap $g = 0.1R$ above a bottom surface. The lengths of the velocity vectors are proportional to the local velocity magnitudes. The velocity vectors are colored by the out-of-plane or $z$-component of the velocity $w$.

The effect of cylinder aspect ratio (defined as $\eta = l/2R$) on the three-dimensionality of the global flow field generated by a rotating horizontally-laid cylinder is illustrated in Fig. 11 by plotting a representative instantaneous streamline passing through a point near the longitudinal axis of the cylinder slightly above the $z = 0$ plane, for aspect ratios $\eta = 0.1$, 0.5 and 1.5. Note that because of the way $l$ and $R$ are defined for this robot (see Fig. 1(b)), low values of the aspect ratio correspond to long and slender cylinders. We see that in general, the streamline can bend or fold significantly due to the cumulative effect of the small $z$-components of the flow velocities. The smaller the $\eta$, the greater the bending of the streamline. For $\eta = 1.5$, the bending is negligible and the streamline stays more or less on the $z = 0$ plane. For $\eta = 0.1$, we see a strikingly large amount of upward bending of the streamline. In this case, the maximum excursion of the streamline above the $z = 0$ plane is on the order of the length ($2R$) of the cylinder. Thus, the three-dimensionality of the global flow field induced by the horizontal cylinder increases dramatically with reduction in the aspect ratio.

Finally, the effect of the size of gap $g$ on the on the three-dimensionality of the global flow field is illustrated in Fig. 12, by looking at the bending of the streamline seeded as above. First of all, we note that even when there is no bottom surface, the streamline bends. The bottom surface enhances the streamline bending for $g/R \leq 1.5$ (the threshold value is between 1.5 and 5). Thus, the three-dimensionality of the global flow field is enhanced by reducing the gap between the robot and the bottom surface, as expected.

In summary, of the four robot shapes studied, a horizontally-laid cylindrical robot, one with low aspect ratio and a small gap, generates the most complicated flow with intricate temporal oscillations and substantial three-dimensionality. It is indeed noteworthy that a simple robot shape like a cylinder executing a simple motion can induce such complex flow patterns. Such highly non-uniform or unsteady flows could be desired for applications such as micromixing. In particular, the large amount of folding or bending of streamlines would increase the "contact area" between different fluids inside the flow fields, and hence improve the mixing efficiency.

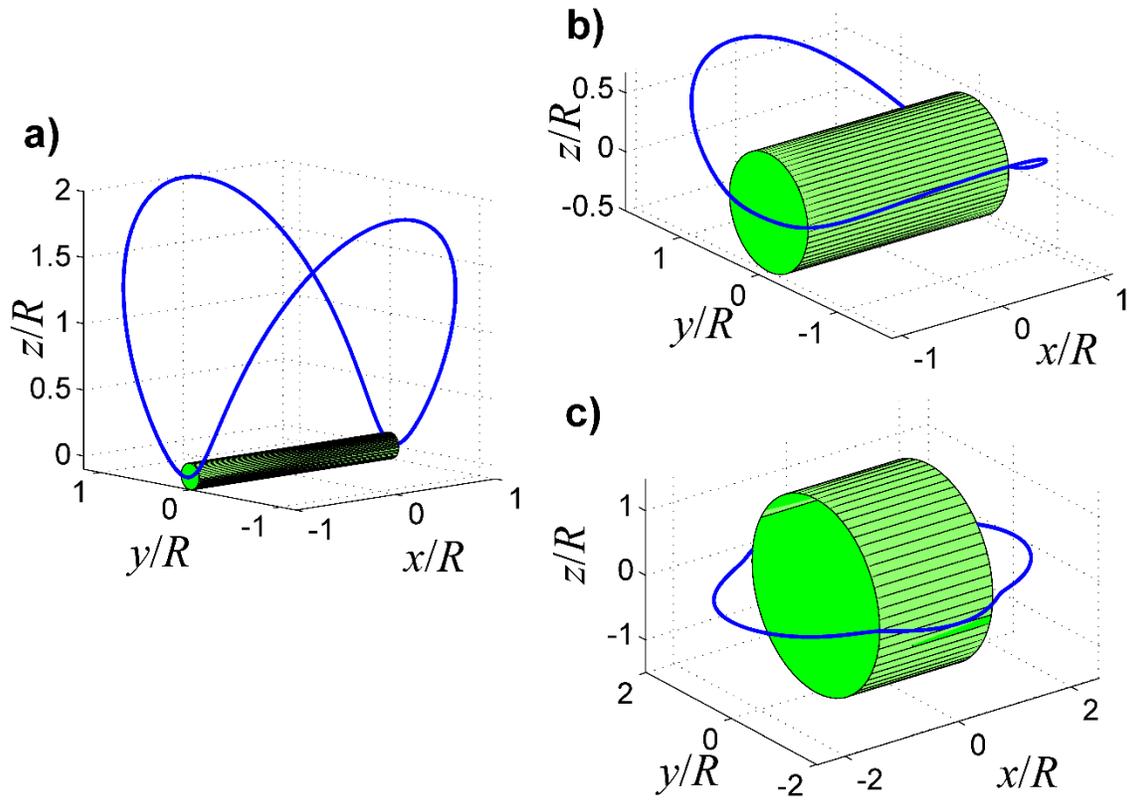

FIG. 11 Instantaneous streamlines of the induced flows around a horizontally-laid cylindrical robot rotating about $z$-axis $g = 0.1R$ above a bottom surface with varying aspect ratios $\eta$ (= $l/2R$). a) $\eta = 0.1$; b) $\eta = 0.5$; c) $\eta = 1.5$. In all three cases, the streamlines are seeded at points on $x$-axis on $z = 0$ plane.

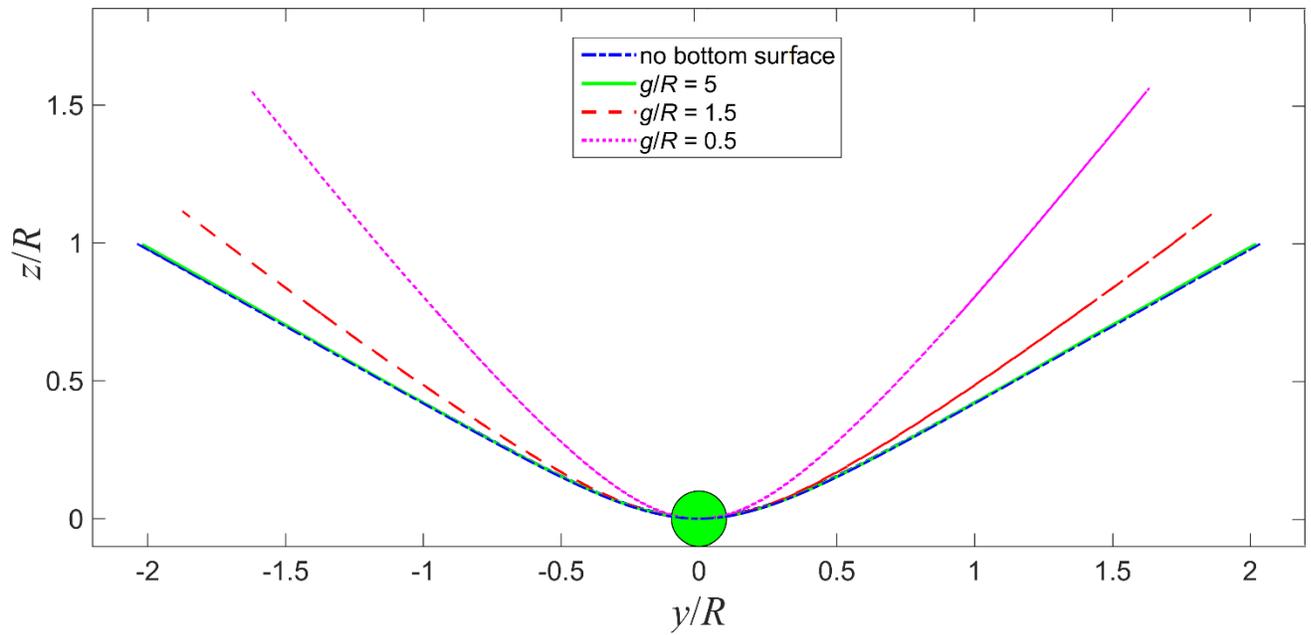

FIG. 12 Streamlines of the induced flows around a horizontally-laid cylindrical robots with aspect ratio $\eta = 0.1$ rotating about $z$-axis above a bottom surface with varying gap size.

## F. Driving torque and power consumption

In previous sections, we mainly focused on characterizing the flow fields induced by rotating microrobots. However, the torque and power input required to actuate these microrobots to generate the rotational flows is an important design consideration to minimize the required actuation power (e.g., remote magnetic actuation), and takes steps towards their future energy-efficient applications. Therefore, in this section, we try to look at the torque/power required for rotating a microrobot in a viscous liquid. Noticing that the power consumption of a rotating robot is given by $P = T\omega$, at a given rotation speed $\omega$, comparing the actuation or driving torques $T$ is sufficient to judge the relative power requirements of different microrobots. When a robot rotates in steady state, it experiences zero net torque, and hence, $T_v + T = 0$, where $T_v$ is the resistive viscous torque applied on the robot by the surrounding liquid due to its rotation. Therefore, we can readily determine the magnitude of $T$ from the value of $T_v$ which can be directly calculated from our Stokes flow BEM code by integrating the surface fluid stresses over the entire robot.

We first look at how the robot shape affects the required driving torque. In the following analysis, all the torque values are normalized by the viscous torque on a rotating sphere in an unbounded fluid,[23,24] $T_0 = 8\pi\mu\omega R^3$. We choose $\mu = 0.001$ Pa·s, $\omega = 314.16$ rad/s, and $R = 10$ μm, as these are the values used in our simulations. The same setup as in Section IVA is used, and the resulting values of $T_v$ for the four different robot shapes are listed in Table 1. The accuracy of the derived torques is verified by comparing the torque calculated by our BEM code (1.1191) for the spherical robot rotating at a $g = 0.1R$ to that derived in the literature (1.1186), as the relative error is less than 0.05%.[23,24]

First, we see that the presence of the bottom surface increases the torques required to rotate the robots. The driving torque for a sphere located $R/10$ above a surface is about 12% greater than a sphere in unbounded fluid. Smaller gaps $g$ requires larger driving torques. Next, we observe that the spherical shape requires the least driving torque to rotate the robot, while the horizontally-laid cylinder, even with aspect ratio 1, requires the highest, equal to about 234% of that for the spherical shape. Longer aspect ratios require even more torque, as discussed later. Although it provides the most uniform induced flow field, the upright cylindrical shape also requires a driving torque significantly larger (by about 217%) than the sphere. This means that given a fixed power input, a spherical robot shape can achieve the highest rotation speed among the four shapes, resulting in a rotational flow field with the highest maximum flow velocity. In addition, based on the results from Fig. 5, a larger maximum flow velocity also means a larger distance affected by the induced rotational flows, which yields in a larger operation range for applications such as micromanipulation. Therefore, a spherical robot shape may be more optimal than the other three shapes for applications that require a relatively steady field in a large operation range, at a fixed power consumption rating.

TABLE 1. Normalized viscous torques acting on the rotating microrobot with different shapes at a fixed rotation speed. The values are normalized by the viscous torque on a sphere rotating at the same speed in an unbounded fluid.

| *Shape* | Sphere | Upright Cylinder | Horizontally-Laid Cylinder | Star |
|---|---|---|---|---|
| *Torque* | 1.12 | 2.43 | 2.62 | 1.63 |

In Section IVD, we analyzed the influence of aspect ratio ($\eta = l/2R$) of the horizontally-laid cylindrical robot on the induced rotational flow field. Here, we also explore how the aspect ratio affects the required driving torque at a given gap $g$. The same setup as in Section IVD is used for this study. However, for this analysis, we change the aspect ratio by using two different approaches: varying $l$ while keeping $R$ constant, and varying $R$ while keeping $l$ constant. To eliminate the effect of gap size $g$, we maintain a constant ratio $g/l = 0.1$ for all simulations. Interestingly, we find that a different non-dimensionalization of the torque is needed for these two sets of results to fall on a single curve, as given below:

$$T_v^* = \frac{T_v}{\mu \omega l S_{robot}} \quad (7)$$

where $S_{robot}$ is the surface area of the robot. The variation of the above non-dimensional torque $T_v^*$ with aspect ratio $\eta$ is shown in Fig. 13. We see that as the aspect ratio is decreased below unity, i.e., as the robots become longer and more slender, the non-dimensionalized torque increases rapidly. However, for aspect ratios above unity, the torque remains essentially constant. To gain more insight of this phenomenon, we conduct a qualitative analysis here: The total surface area of a cylindrical robot can be expressed as $S_{robot} = S_{cap} + S_{side} = \frac{\pi}{2}l^2 + 2\pi l R = \pi l^2(\frac{1}{2} + \frac{1}{\eta})$. Then we proceed the analysis with two cases:

1. When $\eta \ll 1$, the robot is long and slender, so $S_{robot} \approx S_{side} = 2\pi lR$. It is valid to assume that the main contribution to the torque is from the side of the robot. The linear speed $dv$ of a small segment of the robot $dr$ at distant $r$ is $\omega r$, with a surface area of $\pi l \cdot dr$. The local fluid drag on such segment, under Stokes flow assumption, is approximately linearly proportional to its speed and surface area, and hence we have $df \sim dS \cdot dv \sim \omega rl \cdot dr$, and hence the local torque around the rotation axis can be estimated as $dT_v = df \cdot r \sim \omega l r^2 dr$. The total torque can then be obtain by integral over the hole length as $T_v = \int_{-R}^{R} dT_v \sim \omega lR^3$. Therefore, the non-dimensionalized torque would be $T^* = \frac{T_v}{\mu \omega S_{robot} l} \sim \frac{lR^3}{lR \cdot l} = \frac{R^2}{l} = l\eta^{-2}$. It can be seen that the non-dimentionalized torque decreases faster than linearly with aspect ratio at small $\eta$.

2. When $\eta \gg 1$, the robot is short and flat, so it is valid to assume that the main contribution to the torque is from the caps of the robot. A vertical strip with width $dh$ at a distance $h$ from the center of the cap has a surface area of $\sqrt{\frac{l^2}{2} - h^2} \cdot dh$. The linear speed of this strip is $\omega\sqrt{h^2 + R^2}$. The local fluid drag, similar to the previous case, can be approximated as $df \sim dS \cdot dv \sim \omega\sqrt{h^2 + R^2}\sqrt{\frac{l^2}{2} - h^2} \cdot dh$, and hence the local torque, assuming small $R$, is $dT_v \approx df \cdot h \sim \omega(h^2 + R^2)\sqrt{\frac{l^2}{2} - h^2} \cdot dh$. The total torque would be $T_v = \int_{-l}^{l} dT_v \sim \omega l^2(l^2 + R^2)$. Therefore, the non-dimensionalized torque would be $T^* = \frac{T_v}{\mu \omega S_{robot} l} \sim \frac{l^2(l^2 + R^2)}{l^2(\frac{1}{2} + \frac{1}{\eta}) \cdot l} = \frac{l(1 + \frac{1}{4\eta^2})}{(\frac{1}{2} + \frac{1}{\eta})}$. It can be seen that the non-dimensionalized torque remains relatively constant as aspect ratio changes at large $\eta$ if only $R$ is changing.

The above qualitative analysis preliminarily explains the relationship between aspect ratio and the torque normalized by surface area and $l$. We reiterate that only with the above normalization, is it possible to use a single trend to represent the results of the two sets of simulations, allowing us to predict how the driving torque changes with aspect ratio for a horizontally-laid cylindrical robot, regardless of how the change in aspect ratio is achieved.

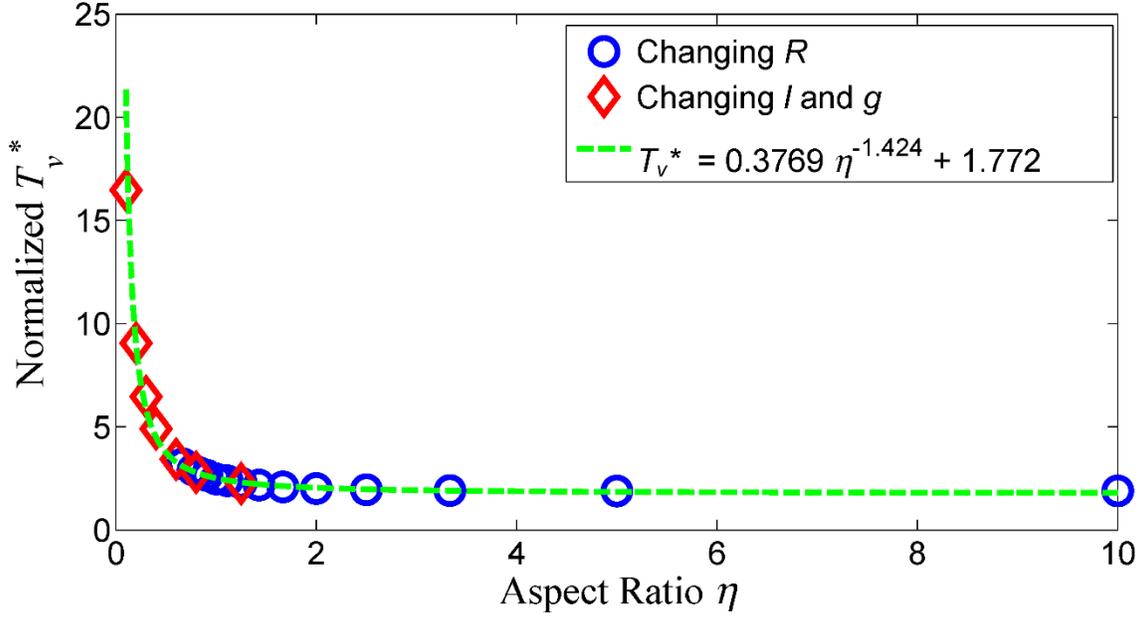

FIG. 13 Influence of aspect ratio of a horizontally-laid cylindrical robot on the viscous torque it experiences.

## V. CONCLUSIONS

In this study, we performed numerical simulations to investigate the rotational flows induced by a microrobot rotating above an underlying surface. We used a custom parallelized solver to numerically solve the Stokes flow problems using the boundary element method. A systematic mesh-independence analysis was carried out to quantify and minimize discretization errors. Validation of the solver's capability to predict steady as well as slightly unsteady creeping flows was done by comparing the results with the predictions from a well-known, commercially available, finite element analysis package. . Four simple robot shapes were studied: a sphere, an upright cylinder, a horizontally-laid cylinder, and a five-pointed star. The flows induced by sphere and upright cylinder are time-invariant and attractive for applications such as micromanipulation, while those due to the horizontally-laid cylinder and five-pointed star are unsteady and potentially useful for applications such as micromixing. The following are our main findings regarding the characteristics of these flows:

1. The flow by generated upright cylinder was the strongest flows and persisted for the greatest distances in both the radial and the vertical directions.

2. Nearby surfaces (bottom substrates and sidewalls) were found to have a significant retarding effect on the induced flows. This effect was quantified, and found to be have significant impact only in their vicinity (within 1-2 robot radii).

3. Of the unsteady flow patterns generated by the two shapes that are not axisymmetric about the rotation axis, the flow pattern of the horizontal cylinder is more intricate than the star, with a comparatively large contribution from higher harmonics. The amplitudes of the unsteady fluctuations diminish with increasing distance from the robots.

4. The flows induced by the horizontally-laid cylinder can be substantially three-dimensional, particularly for long and slender cylinders rotating above a bottom surface with small gaps. It is postulated that this three-dimensionality is due to the curved cylindrical surface, and enhanced by the presence of the bottom surface. This feature could improve the robot's performance in micromixing.

5. In terms of driving force and power consumption requirements, the upright cylindrical robot is the most demanding. In terms of the induced flow strength at a given power consumption rating, the sphere is optimal. Presence of nearby surfaces such as a bottom wall increase driving torque power consumption.

This study provides many insights relevant for optimal designs of microrobots[25] in applications with varying requirements for induced flow fields, driving torque and power consumption.

## ACKNOWLEDGMENTS

M.S. was supported by the NSF National Robotics Initiative Program (NRI-1317477). Access to the Blacklight supercomputer at the Pittsburgh Supercomputing Center was made possible via NSF XSEDE Grant ENG150006.